\documentclass{IPEC2026TEC}
\IEEEoverridecommandlockouts
\usepackage{cite}
\usepackage{amsmath,amssymb,amsfonts}
\usepackage{algorithm}
\usepackage{algpseudocode}
\usepackage[dvipdfmx]{graphicx}
\usepackage{textcomp}
\usepackage{xcolor} % If you are using uplatex or similar tools, please add the [dvipdfmx] option.
\def\BibTeX{{\rm B\kern-.05em{\sc i\kern-.025em b}\kern-.08em
    T\kern-.1667em\lower.7ex\hbox{E}\kern-.125emX}}
\begin{document}

\title{Quantized Probabilistic AI for Gear Fault Diagnosis in Motor Drives}

\author{\IEEEauthorblockN{
    Subham Sahoo\textsuperscript{1}*, 
    Huai Wang\textsuperscript{1}, 
    Frede Blaabjerg\textsuperscript{1}
}
\IEEEauthorblockA{1 Department of Energy, Aalborg University, Denmark}
\IEEEauthorblockA{*email: sssa@energy.aau.dk}
}

\maketitle

\begin{abstract}
 Deploying large artificial intelligence (AI) models in power electronics often demands high computational resources. Driven by the quantization paradigm, this digest proposes a quantization-aware training (QAT) principle to substantially minimize the number of bits required and simultaneously maximize the accuracy of computations in pre-trained AI models. Considering a pre-trained probabilistic Bayesian Neural Network (BNN) for gear fault diagnosis in motor drives as an example, we quantize its weights and activation functions from floating-point FP32 to low-precision INT8 values, which enhances the computational efficiency by a significant margin of 30-45\% (for different model versions) without any compromise in the accuracy and uncertainty estimates. This substantiates a sustainable mechanism of deploying most quantized light-weight AI models into low-cost edge processors for power electronic applications.
\end{abstract}


\section{Introduction}
Artificial intelligence (AI)-driven fault diagnosis in motor drives must run on constrained hardware (on-device, edge microcontrollers) while providing robust decisions under noisy sensors and unseen conditions \cite{yliao,roja}. Bayesian neural networks (BNNs) naturally provide calibrated predictive distributions (epistemic and aleatoric uncertainty) \cite{encon,iem}, which helps quantify diagnostic confidence and improves safety in the face of sensor noise or domain shift. Recent work has shown BNNs applied to gear-fault diagnosis in motor drives deliver interpretable uncertainty estimates and robustness to noisy/unseen data \cite{bnn1}. However, training of probabilistic weights can be quite time-consuming and demand a lot of resources.

While model quantization is widely adopted to reduce computational complexity of AI models, it does not inherently guarantee the preservation of model accuracy or reliability. As shown in Fig. 1(a), direct quantization of a pre-trained BNN from FP32 to INT8 introduces rounding noise that often distorts the learned decision boundaries \cite{quant1, quant2}, thereby compromising classification accuracy and uncertainty calibration. 

To address this limitation, a Quantization-Aware Training (QAT) mechanism is proposed in this paper (see Fig. 1(b)), where quantization effects are simulated during training so that the model can adapt its parameters to the low-bit environment \cite{nagel}. Unlike conventional quantization that treats precision reduction as a post-processing step, QAT jointly optimizes both model weights and quantization thresholds, effectively learning to perceive quantization noise as part of the training domain. As depicted, the proposed method preserves the angular relationships among fault classes and yields well-calibrated, high-confidence fault classification despite reduced bit precision. 
\begin{figure}[t]
   \centering
    \includegraphics[width=0.45\textwidth]{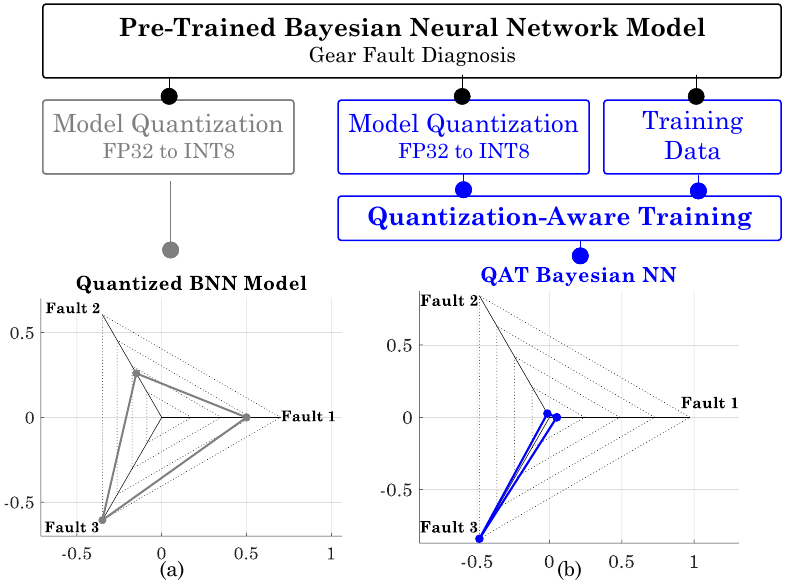}
   \caption{Quantization approaches for probabilistic BNNs: (a) direct model parameter quantization, (b) Quantization-aware training (QAT) \cite{nagel} of the pre-trained model ensuring high classification accuracy. \textbf{It should be noted that FP32 is a single-precision floating computer data format requiring 32 bits in computer memory, whereas INT8 requires 8 bits.}}
    \label{Fig_2}
\end{figure}
Building on this foundation, the key contributions of this work are:
\begin{enumerate}
    \item \textbf{QAT-enabled probabilistic AI for motor drives}: A novel integration of quantization-aware training into Bayesian neural networks, ensuring quantization robustness without loss of predictive uncertainty.
\item \textbf{Empirical validation}: Comprehensive benchmarking using the Gearbox Dynamic Simulator dataset, achieving 30–45\% compute efficiency gain and $\approx$ 4$\times$memory reduction with negligible accuracy degradation and different noise levels.
\item \textbf{Compute reliability trade-off framework}: Analytical mapping of model accuracy, computational cost, and boundary shifts across quantization levels to guide edge hardware deployment.
\end{enumerate}
The proposed QAT-based probabilistic framework is particularly well-suited for deployment in resource-constrained systems. Unlike FP32-based models that typically require GPU-level computational resources, INT8 quantized models can be efficiently executed on embedded processors such as DSPs, ARM Cortex microcontrollers, and FPGA-based accelerators that support integer arithmetic. This transition from floating-point to integer computation enables faster inference due to simplified arithmetic pipelines and reduced memory bandwidth requirements. Furthermore, the reduction in model size, as shown in Fig. 6, allows the deployment of uncertainty-aware AI models within the limited memory budgets of edge devices. This makes the proposed approach highly relevant for real-time fault diagnosis in industrial motor drives, where low latency, reliability, and energy efficiency are critical.
%The inherent fault signatures in motor drive datasets are leveraged to converge with \textit{physics-aware training} of a single readout layer for fine tuning of their embodied probabilistic dependencies. In this manner, this tailored design mechanism not only reduces the computational efforts and dependence on data science preliminaries, but also bridges the knowledge gap between the modeling of AI in power electronics.
\begin{figure}[t]
    \centering
    \includegraphics[scale=0.55]{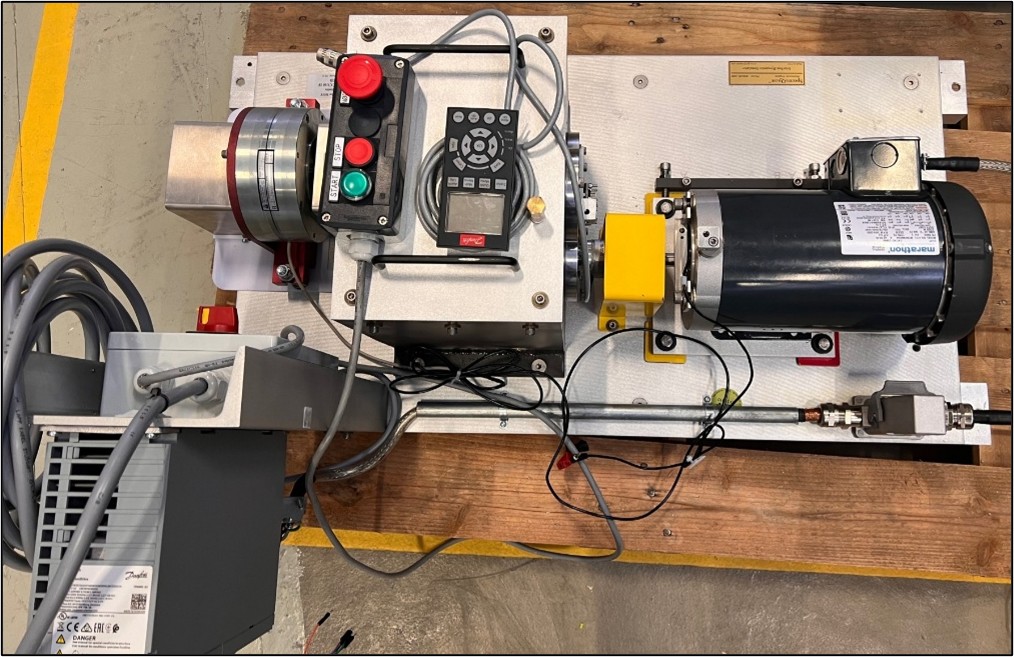}
    \caption{Experimental testbed with SpectraQuest Gearbox Dynamic Simulator and Danfoss VLT Drive FC-103 for collection of data in Table I.}
    \label{Fig_32}
\end{figure}
\begin{figure}[t]
    \centering
    \includegraphics[scale=2.3]{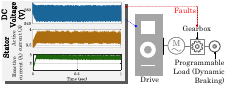}
    \caption{Layout of Gearbox Dynamic Simulator \cite{gds} for collection of different gear fault (re-configurable) data, obtained under different loading conditions.}
    \label{Fig_3}
\end{figure}
\section{System Specifications}
SpectraQuest’s Gearbox Dynamics Simulator (GDS), shown in Fig. 2 and 3, is used to simulate industrial gearboxes and emulate gear faults. More details on this setup can be found in \cite{agni}.
Gear faults, that typically manifest as cracks on the gear or wear and tear of the gear teeth, were pre-configured for a dynamic brake as programmable load for data collection under different conditions. Hence, the gearbox in Fig. 2 and 3 can be easily swapped and reconfigured to the faults in Table I.
\begin{table}[h!]
\centering
\caption{Gear Fault Data Preliminaries}
\begin{tabular}{c|c||c}
\hline
\textbf{Fault Label} & \textbf{Fault} & \textbf{Intrinsic Data} \\
\hline
1 & Missing tooth & Speed  \\
2 & Chipped tooth & Motor torque \\
3 & Root crack & DC link voltage\\
4  & Surface crack & Active stator current\\
5 & Eccentricity & Reactive stator current\\
\hline
\end{tabular}
\end{table}

Two types of sensors were used in this project, \textit{intrinsic} (within the drive) and \textit{extrinsic} (exterior to the drive). The setup in Fig. 2 was modified with a Danfoss VLT Drive FC-103 to provide the intrinsic measurements listed in Table I, each sampled at a frequency of 5 kHz. On the other hand, a pair of extrinsic orthogonally aligned analogue accelerometers ADXL1001 are mounted on the gearbox. A probabilistic Bayesian Neural Network (BNN) was employed to perform gear fault diagnosis using intrinsic measurements in Table I. This pre-trained BNN in \cite{bnn1} was designed to capture uncertainty in both model parameters and fault classification outcome, thereby enhancing diagnostic reliability under noisy and unseen conditions. Its specifications are listed in Table II.
\begin{table}[t!]
\centering
\caption{BNN \cite{bnn1} Preliminaries}
\begin{tabular}{c|c||c}
\hline
\textbf{Parameters} & \textbf{Quantity} & \textbf{Training Data} \\
\hline
Convolutional layers & 5 (2D networks) & Fault 1  \\
Max-pooling & 2$\times$1 & Fault 2 \\
Hidden layers (Neurons) & 3 hidden (128,64,32) & Fault 3\\
Output & 3 & -\\
Optimizer & Stoch. gradient descent & -\\
\hline
\end{tabular}
\end{table}
\begin{figure*}[t]
    \centering
    \includegraphics[scale=3.45]{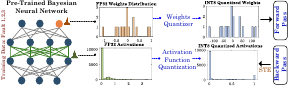}
    \caption{Quantization-aware-training framework \cite{nagel}--Conversion of weights and activation functions in pre-trained BNN in Table II for quantized inferences. In the backward pass for INT8 weights, gradients are passed through as if the rounding operation were an identity function using the \textbf{Straight-Through Estimator} (STE) function \cite{jung}.}
    \label{Fig_3}
\end{figure*}

It should be noted that the QAT-based BNN will be finally tested with unseen data Fault 4 \& 5 in Table I.
\section{Comparing Model Quantization and Light AI}
Lightweight AI methods, such as pruning or architectural compression \cite{light1,light2} minimize the parameter count \( P' < P \) or layer depth \( L' < L \), but they alter the original hypothesis space \( \mathcal{H} \) of the original AI model, leading to a perturbed mapping \( f'(x) \neq f(x) \). In contrast, Quantization-Aware Training (QAT) retains the pre-trained model structure and embeds \textit{quantization} during training, such that:
\begin{equation}
    \hat{w} = Q(w) = \mathrm{round}\!\left(\frac{w}{\Delta}\right)\Delta, 
    \qquad 
    \mathcal{L}_{\text{QAT}} = \mathcal{L}_{\text{data}}\big(f(\hat{w},x),y\big),
\end{equation}
where, \( \Delta \) denotes the quantization step size. The gradients are computed through a \textit{straight-through estimator (STE)} \cite{jung} ensuring:
\begin{equation}
    \nabla_w \mathcal{L}_{\text{QAT}} \approx \nabla_{\hat{w}} \mathcal{L}_{\text{data}},
\end{equation}
which preserves the optimization trajectory of the full-precision network while learning robustness to quantization noise. Hence, QAT acts as a \textit{precision regularizer} that constrains weight dynamics within discrete manifolds without compromising the pretrained model’s success. This is helpful in reducing the compute memory requirements for pre-trained AI models tailormade for power electronic applications, instead of designing a new lightweight model without any theoretical performance guarantees.

The impact of quantization can be systematically analyzed through a trade-off between classification accuracy, uncertainty calibration, and computational cost. Let $\mathrm{A}(b)$, $\mathrm{U}(b)$, and $\mathrm{C}(b)$ denote the accuracy, uncertainty calibration error, and computational cost of the model at bit-width $b$, respectively. The design objective for edge deployment can then be formulated as:
\begin{equation}
b^* = \arg\min_{b} \; \mathrm{C}(b)
\quad \text{subject to} \quad
\mathrm{A}(b) \geq \mathrm{A}_{\min}, \;
\mathrm{U}(b) \leq \mathrm{U}_{\max}
\end{equation}
where, $\mathrm{A}_{\min}$ and $\mathrm{U}_{\max}$ represent application-specific thresholds for acceptable accuracy and uncertainty. This formulation highlights that quantization is not merely a compression tool, but a design parameter that governs the balance between performance and efficiency.

\section{Quantization-Aware Training of Probabilistic AI for Motor Drives}
Conventional post-training quantization from FP32 to INT8 rarely preserves accuracy or diagnostic confidence, as the network was never optimized for quantization noise. To overcome this, \textit{Quantization-Aware Training (QAT)} \cite{nagel} embeds the quantization process directly within training so that the model learns to operate natively in low precision.

\subsection{Quantization-aware training} As illustrated in Fig.~3, the workflow begins with a \textit{pre-trained Bayesian Neural Network (BNN)} trained on gearbox fault data (Fault~1, 2 \& 3), both the FP32 \textit{weights} and \textit{activations} are quantized to INT8 through learned quantizers, producing low-bit counterparts that approximate the original full-precision mapping,
\begin{equation}
    \| f_{\text{FP32}}(x) - f_{\text{INT8}}(x) \|_2 \leq \epsilon
\end{equation}
where, $\epsilon$ denotes the learned quantization tolerance. In the \textbf{backward pass}, gradients flow through the non-differentiable rounding operation using the \textit{Straight-Through Estimator (STE)} \cite{jung}:
\begin{equation}
    \frac{\partial}{\partial x}\text{round}(x) \approx 1,
\end{equation}
allowing standard gradient-based optimization to continue while incorporating quantization effects. Through this dual-stage process, the BNN gradually adapts its parameters to compensate for quantization noise, yielding INT8 models that retain predictive accuracy and calibrated uncertainty. Hence, QAT outlined in Algorithm I enables a \textit{noise-aware learning mechanism}, ideally suited for reliable, low-power gear-fault diagnosis in motor drives.

In a Bayesian Neural Network (BNN), each model parameter is represented as a probability distribution rather than a deterministic scalar. Given training data $\mathcal{D} = \{x_i, y_i\}$, the posterior over weights is expressed as:
\begin{equation}
    p(\mathbf{w}|\mathcal{D}) = \frac{p(\mathcal{D}|\mathbf{w})\,p(\mathbf{w})}{p(\mathcal{D})},
\end{equation}
where, $p(\mathcal{D}|\mathbf{w})$ denotes the \textit{likelihood}, and $p(\mathbf{w})$ represents the \textit{prior distribution} of the weights \cite{bnn1}. 
For inference, the predictive distribution over the target $y^*$ for an unseen input $x^*$ is computed as:
\begin{equation}
    p(y^*|x^*, \mathcal{D}) = \int p(y^*|x^*, \mathbf{w}) \, p(\mathbf{w}|\mathcal{D}) \, d\mathbf{w}.
\end{equation}

\subsection{Quantization of Bayesian weights}  
During quantization-aware training (QAT) in Fig. 4, each stochastic weight sample $w$ drawn from the learned posterior $p(\mathbf{w}|\mathcal{D})$ is quantized to a finite set of discrete levels. 
Let the quantizer $\mathcal{Q}(\cdot)$ map a continuous value to the nearest quantized representation:
\begin{equation}
    \mathcal{Q}(w) = \text{clip}\!\left(\text{round}\!\left(\frac{w}{S}\right), -2^{b-1}, 2^{b-1}-1 \right) S
\end{equation}
where, $S = \frac{\beta-\alpha}{2^b - 1}$ is the quantization scale factor, $b$ is the bit width, and $[\alpha,\beta]$ is the clipping range for the corresponding tensor. 
In this way, the quantized posterior can be approximated as:
\begin{equation}
    \tilde{p}(\mathbf{w}_q|\mathcal{D}) \approx \prod_i \delta(\mathbf{w}_q^{(i)} - \mathcal{Q}(\mathbf{w}^{(i)}))
\end{equation}
where, $\delta(\cdot)$ denotes the Dirac delta function representing quantization to discrete levels.  

\subsection{Quantization of activations}  
Similarly, activations $h_i = f_i(h_{i-1}; \mathbf{w}_i)$ are quantized after each layer to simulate integer arithmetic during forward propagation:
\begin{equation}
    h_{i,q} = \mathcal{Q}(h_i) = \text{clip}\!\left(\text{round}\!\left(\frac{h_i}{S_h}\right), -2^{b-1}, 2^{b-1}-1 \right) S_h
\end{equation}
where, $S_h$ is the scale factor for the activations. This ensures that both the weights and intermediate feature maps are represented in low precision during training.

%\textbf{Loss function adaptation.}  
%For the probabilistic BNN, the total loss incorporates both the negative log-likelihood and a Kullback–Leibler (KL) divergence regularization term. During QAT, this becomes:
%\begin{equation}
%    \mathcal{L}_{\text{QAT}} = \mathbb{E}_{\tilde{p}(\mathbf{w}_q|\mathcal{D})}\!\left[-\log p(\mathcal{D}|\mathbf{w}_q)\right] 
%    + \text{KL}\!\left(\tilde{p}(\mathbf{w}_q|\mathcal{D}) \, \| \, p(\mathbf{w})\right),
%\end{equation}
%which enforces both predictive accuracy under quantization and posterior consistency with the original Bayesian prior. 

%\textbf{Gradient approximation using STE.}  
%Since the quantization operation is non-differentiable, QAT employs the \textit{Straight-Through Estimator (STE)} to approximate gradients during backpropagation:
%\begin{equation}
%    \frac{\partial \mathcal{Q}(x)}{\partial x} \approx 1, \quad \text{for } |x| < \beta.
%\end{equation}
%This allows the optimizer to update full-precision weights while simulating quantized inference behavior. Over successive epochs, the BNN parameters adapt such that:
%\begin{equation}
%    \| f_{\text{FP32}}(x) - f_{\text{INT8}}(x) \|_2 \rightarrow \min_\theta,
%\end{equation}
%minimizing the quantization-induced deviation between full- and low-precision outputs.

\begin{algorithm}[!t]
\caption{QAT workflow for BNNs}
\begin{algorithmic}[1]
\State \textbf{Input:} Pre-trained BNN $\mathcal{L}(\theta)$, training data $\mathcal{D} = \{x_i, y_i\}$
\State Initialize quantization parameters: bit width $b$, clipping range $[\alpha, \beta]$
\For{each training iteration}
    \State Quantize weights: $w_q = \text{Quantize}(w, b, [\alpha, \beta])$
    \State Quantize activations: $h_q = \text{Quantize}(h, b, [\alpha, \beta])$
    \State Perform \textbf{forward pass} with $(w_q, h_q)$ to compute loss $\mathcal{L}$
    \State Compute gradients using \textbf{STE}: $\frac{\partial}{\partial x}\text{round}(x) \approx 1$
    \State Update weights: $\theta \leftarrow \theta - \eta \nabla_\theta \mathcal{L}$
\EndFor
\State \textbf{Output:} Quantization-aware trained model $\mathcal{L}_{\text{QAT}}(\theta_q)$
\end{algorithmic}
\end{algorithm}
In the context of gear fault diagnosis, this QAT-enabled BNN effectively learns to internalize quantization noise as part of its uncertainty model. 
\subsection{Interpretation of quantization in probabilistic models}  
Quantization in probabilistic neural networks can be interpreted as a structured form of noise injection that interacts naturally with the uncertainty modeling capability of Bayesian frameworks. 

In conventional deterministic networks, quantization introduces rounding errors that distort decision boundaries. However, in Bayesian Neural Networks (BNNs), where weights are represented as probability distributions, such perturbations can be absorbed within the learned posterior. From this perspective, quantization acts as a constraint that projects continuous weight distributions onto discrete manifolds, effectively regularizing the model. This allows the network to internalize quantization noise as part of its uncertainty representation, rather than treating it as an external disturbance. As a result, the predictive distributions remain calibrated even under reduced precision, explaining the observed preservation of both classification accuracy and uncertainty estimates in the QAT-based framework.
\subsection{Synergy of quantization to physical characteristics of power electronic systems}

The discrete nature of quantization aligns well with the operational characteristics of power electronic systems. In practical motor drives, signals are inherently digitized through analog-to-digital converters and influenced by switching dynamics, resulting in discretized and noisy measurements. From this perspective, quantization-aware training can be interpreted as embedding these physical constraints directly into the learning process. By incorporating quantization effects during training, the model becomes inherently robust to signal discretization and measurement noise, which are intrinsic to power electronic environments. This further strengthens the physical consistency of the proposed framework, bridging the gap between digital control systems and probabilistic AI modeling.
\begin{figure}[t]
    \centering
    \includegraphics[width=0.38\textwidth]{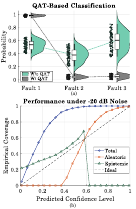}
    \caption{Performance evaluation--(a) Comparative evaluation of pre-trained BNN with and without QAT, (b) Performance of QAT-based BNN under -20 dB noise--rise in aleatoric uncertainty (accounting for noise) affects the confidence level.}
    \label{Fig_3}
\end{figure}
\begin{figure}[t]
    \centering
    \includegraphics[width=0.4\textwidth]{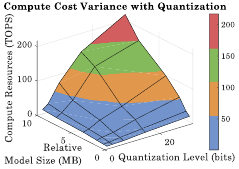}
    \caption{Compute cost variance of the QAT-based BNN with respect to different quantization levels.}
    \label{Fig_3}
\end{figure}
\begin{figure}[t]
    \centering
    \includegraphics[width=0.45\textwidth]{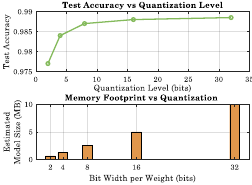}
    \caption{Test accuracy and memory footprint of the QAT-based BNN model at discrete quantization levels--measured using \cite{hugging}.}
    \label{Fig_6}
\end{figure}
\begin{figure}[t]
    \centering
    \includegraphics[width=0.35\textwidth]{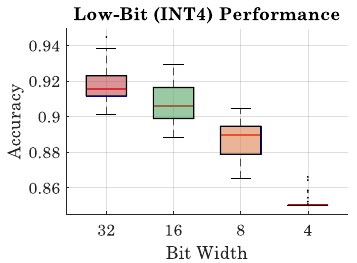}
    \caption{Performance under extreme low-bit quantization--classification accuracy shows a gradual degradation at INT4 precision, which translates into application-dependent decision thresholds for design engineers.}
    \label{Fig_7}
\end{figure}
\section{Performance Evaluation}
To investigate the performance, the model training and inference of QAT-based BNN is carried out on a NVIDIA \textit{RTX 4050} GPU hardware with the memory as well as compute resources estimated in \textit{Hugging Face} \cite{hugging} environment.

The classification behavior of QAT-based BNN under seen data conditions is shown in Fig.~4(a). The box plots compare the predictive probability distributions obtained \textit{with} and \textit{without} QAT for the three representative gear faults. Without QAT, the model exhibits wider probability spreads and lower mean confidence across all fault classes, indicating a high degree of epistemic uncertainty even on known operating conditions. This suggests that direct post-training quantization distorts the BNN’s learned decision boundaries, resulting in less calibrated fault predictions.

In contrast, the QAT-based BNN maintains a sharper and more compact probability distribution for each fault class. The confidence levels are significantly higher, and the variance within class predictions is markedly reduced. This reflects that the QAT process successfully adapts the model parameters to the quantization-induced perturbations during training, thereby aligning the quantized INT8 inference behavior with the original FP32 network. Consequently, the QAT-based BNN produces fault classifications that are both \textit{accurate} and \textit{confidence-calibrated}, ensuring that probabilistic decisions remain consistent with those of the full-precision model.

Fig.~4(b) further investigates model robustness by introducing a $-20$~dB Gaussian noise to the input data. The empirical coverage plot separates the total uncertainty into its \textit{aleatoric} (data-dependent) and \textit{epistemic} (model-dependent) components. Even under heavy noise, the QAT-based BNN preserves high total calibration fidelity. The decomposition further shows that aleatoric uncertainty dominates in high-confidence regions—consistent with the physical expectation that sensor noise primarily affects measurement reliability—while epistemic uncertainty increases in low-confidence regions, reflecting the model’s uncertainty about unseen conditions.  These observations confirm that the proposed QAT framework not only compresses the Bayesian model but also enhances its ability to maintain reliable uncertainty estimates in noisy motor-drive environments. Hence, quantization in this form serves as both a computational optimization and an implicit regularization mechanism, promoting robust and interpretable fault diagnosis in motor drives.

The computational cost variance across different quantization levels is illustrated in Fig.~5. The 3D surface map relates the model size (in MB), quantization level (bit width), and the corresponding compute resources measured in tera-operations per second (TOPS). As the bit-width decreases from FP32 to INT8 and below, both the model size and required computational resources exhibit a near-monotonic reduction. This trend highlights that quantization enables a predictable scaling of computational complexity, primarily due to the transition from floating-point to integer arithmetic, which significantly reduces memory bandwidth and arithmetic overhead. Notably, the reduction in compute cost follows an approximately smooth and near-linear trend with decreasing bit-width, indicating that quantization can be treated as a controllable design parameter rather than an ad-hoc compression step. Hence, designers can systematically select an appropriate bit-width to balance computational efficiency and model performance, without encountering abrupt degradation in accuracy or reliability.

As the quantization level decreases toward lower bit widths (e.g., 8-bit and below), the computational cost exhibits a monotonic reduction. This is evident from the downward slope of the surface in Fig. 6, which reflects the transition from high-precision arithmetic to more compact integer operations. At INT8 precision, the compute resource requirement drops by nearly 30--45\%, accompanied by a corresponding reduction in memory footprint of approximately four times, as validated through profiling on the \textit{NVIDIA RTX~4050} GPU. 
%Overall, the mapping in Fig.~\ref{Fig_Compute_Cost} demonstrates that the proposed QAT-based Bayesian Neural Network enables scalable compute-resource allocation across varying hardware capacities, offering a pathway for sustainable and deployable AI within power-electronic converters and industrial motor-drive platforms.
Further in Fig. 7, minimal accuracy loss is observed as quantization level increases. Nevertheless, even at low bit-widths (4-8 bits), the test accuracy remains above 98.5\%, which indicates that QAT preserves model performance of pre-trained BNN. Moving from 32-bit to 8-bit reduces estimated model size from 10 MB to $<$2.5 MB, which highlights around 4 times reduction in memory cost, making them viable for edge deployment. The trends observed in Fig. 6-7 align with the trade-off formulation in (3), where INT8 represents a practical operating point balancing computational cost and accuracy.
%Finally in Fig. 3(d), the decision boundaries between FP32 and quantized INT8 BNN shows minimal difference, with a narrow band of divergence. This indicates that the quantized model's decision boundaries closely follow the full-precision BNN, without distorting model uncertainty or predictive confidence.

The results in Fig. 8 under low-bit quantization (INT4) indicate that while classification accuracy remains consistently high for INT8 and above, a noticeable degradation emerges at extreme low precision. This behavior suggests that the decision boundaries of the model become increasingly compressed and susceptible to misclassification as the representational resolution decreases. From a practical standpoint, this motivates the introduction of adaptive decision thresholds to maintain reliable operation under constrained precision. Specifically, instead of relying on fixed maximum-probability classification, a minimum confidence threshold can be imposed, below which decisions are either rejected or flagged for further evaluation. This allows the system to trade-off between classification coverage and reliability, where lower bit-widths require higher rejection rates to maintain acceptable diagnostic accuracy.

Despite its advantages, the proposed QAT-based probabilistic framework has certain limitations:
\begin{itemize}
    \item aggressive quantization to very low bit-widths (e.g., below 4 bits) may lead to degradation in both accuracy and uncertainty calibration due to excessive discretization of weight distributions.
    \item the effectiveness of quantization-aware training depends on the availability of representative training data, as the model must learn to adapt to quantization noise during training.
\end{itemize}
These aspects highlight potential directions for future research toward improving robustness under extreme quantization constraints.

\section{Conclusion}
This paper demonstrates that quantization-aware training (QAT) enables Bayesian neural networks to operate efficiently in low-precision regimes while preserving both predictive accuracy and uncertainty calibration. Experimental results show that INT8 quantization achieves a 30–45\% reduction in computational cost and approximately 4× reduction in memory footprint (from 10 MB to $<$2.5 MB), without compromising diagnostic performance. Notably, the classification accuracy remains above 98.5\% even at low bit-widths (4–8 bits), confirming that the proposed approach retains the discriminative capability of the original FP32 model.

Furthermore, under severe noise conditions (-20 dB), the QAT-based model maintains well-calibrated predictive uncertainty, with a clear separation between aleatoric and epistemic components. Compared to direct post-training quantization, the proposed method significantly reduces the spread of predictive distributions for seen faults, indicating improved confidence calibration and robustness. These results collectively demonstrate that QAT not only compresses the Bayesian model but also acts as an implicit regularization mechanism, enabling reliable and interpretable fault diagnosis under constrained precision.

Hence, the proposed framework establishes a practical pathway for deploying uncertainty-aware AI models on resource-limited edge hardware in power electronic systems, without compromising its performance.
%\section*{SUBMISSION OF THE EXTENDED SUMMARY FOR IPEC}
%\textcolor{red}{An extended summary of Regular Session(RS), Organized Session(OS) and Industry Technology Session(ITS) describing work not previously published or presented must be electronically submitted in a PDF file through the conference website no later than October 30th, 2025.} The submitted extended summary will be reviewed via a peer review process in order to ensure the highest technical quality of the conference. The extended summary should clearly define the salient concepts and novel features of the work. Be sure to mention past or previous works to distinguish your originality from them. The extended summary should be up to 4 pages except Reference, and detailed instructions will be shown on the IPEC-Nagasaki official website, \url{  https://www.ipec2026.org}. Authors will receive notification of acceptance by e-mail on or before January 30th, 2026. Proceedings will be published on IEEE Xplore. Detailed instructions are available on the IPEC-Nagasaki 2026 website http://www.ipec2026.org.

\vspace{12pt}

\end{document}